\documentclass[nofootinbib,amsmath,amssymb,aps,floatfix,notitlepage]{revtex4-1}
\pdfoutput=1

\usepackage{graphicx}% Include figure files

\usepackage[utf8]{inputenc}
\usepackage{mwe}
\usepackage{listings}
\usepackage{geometry}

\usepackage{amssymb}
\usepackage{amsmath}
\usepackage{epstopdf}
\usepackage{xcolor}
\usepackage{mathtools}
\usepackage{comment}

\def\hhref#1{\href{http://arxiv.org/abs/#1}{arXiv:#1}} 
\usepackage{hyperref}

\begin{document}  

\title{Resurgent Trans-Series for Generalized Hastings-McLeod Solutions}

\author{Nikko J. Cleri and Gerald V. Dunne}

\affiliation{Department of Physics, University of Connecticut, Storrs, CT 06269-3046, USA}

%\date{\today}

\begin{abstract}
We show that the physical Hastings-McLeod solution of the integrable Painlev\'e II equation generalizes in a natural way to a class of non-integrable equations, in a way that preserves many of the significant qualitative properties. We derive the  trans-series structure of these generalized solutions, demonstrating that integrability is not essential for the resurgent asymptotic properties of the solutions.

\end{abstract}

%\date{\today}

\maketitle

\section{Introduction}
\label{sec:intro}

Painlev\'e equations arise in a wide range of applications in physics, as a class of nonlinear special functions \cite{clarkson,mccoy,bender-book,mason,forrester,tracy-widom}. They are integrable in the sense that their moveable singularities are only poles \cite{ince,nist-painleve}. They have also been widely studied in the context of resurgent asymptotics \cite{ecalle,costin-odes,costin,marino,Garoufalidis:2010ya, Aniceto:2011nu,marino-book,ahmed,dunne}, illustrating the deep connections between perturbative and non-perturbative expansions, for both asymptotic and convergent expansions. For example, the 
Painlev\'e II equation describes the double-scaling limit of the Gross-Witten-Wadia unitary matrix model \cite{gw,wadia,neuberger,marino,ahmed}, which has a third order phase transition in the infinite $N$ limit, the immediate vicinity of which is characterized by the Hastings-McLeod solution of the Painlev\'e II equation \cite{rosales,hastings,bothner}. As nonlinear equations, the Painlev\'e equations exhibit exponential sensitivity to boundary conditions, leading to intricate patterns of separatrices \cite{nist-painleve,rosales,hastings,bender}. In this paper, we investigate these phenomena of resurgence and exponential sensitivity in {\it non-integrable} deformations of the Painlev\'e II equation, using a combination of numerical and analytic methods. We find the full trans-series structure, and 
confirm the general result that the resurgent structure of nonlinear differential equations does not rely on integrability  \cite{costin-odes,costin}.

The Painlev\'e II equation reads
\begin{eqnarray}
y''(x)=2 y^3(x)+x\, y(x)+\alpha
\label{eq:p2}
\end{eqnarray}
With vanishing parameter ($\alpha=0$), this equation has a unique real solution, known as the Hastings-McLeod solution \cite{hastings}, satisfying the following boundary conditions at $x=\pm\infty$:
\begin{eqnarray}
%\lim_{x\to\infty} 
y(x) &\sim& {\rm Ai}(x) \qquad, \qquad x\to+\infty
\label{eq:p2bc1} \\
%\lim_{x\to-\infty} 
y(x) &\sim& \left(\frac{-x}{2}\right)^{\frac{1}{2}}  \qquad, \quad x\to-\infty
\label{eq:p2bc2}
\end{eqnarray}
The power-law behavior as $x\to-\infty$ follows from a balance of the $y^3$ and $x\, y$ terms (when $\alpha=0$).
The Airy behavior as $x\to+\infty$ follows  from the linearization of the differential equation (\ref{eq:p2}) (with $\alpha=0$), but this linearization does not fix the coefficient of the Airy function solution: $y(x)\sim k\, {\rm Ai}(x)$.
The non-trivial result  for the Hastings-McLeod solution is that the coefficient of the Airy function in (\ref{eq:p2bc1}) must be exactly 1 in order to match smoothly with the power-law behavior as $x\to-\infty$.  If $k>1$ the solution diverges for $x<0$, while if $k<1$ the solution oscillates for $x<0$ \cite{rosales,hastings}. See Figure \ref{fig:k3} and \cite{nist-painleve} for graphics. The full asymptotics of this Hastings-McLeod solution is expressed as a trans-series, including both (``perturbative'') power-series terms as well as exponentially-suppressed (``non-perturbative'') terms, whose structures are entwined by resurgent relations \cite{costin,marino,ahmed}. The form of this trans-series is very different as $x\to+\infty$ and as $x\to -\infty$, changing its structure 
as we cross the phase transition at $x=0$ \cite{marino,ahmed}; this is an explicit realization of the physical phenomenon of instanton condensation \cite{neuberger}, in which all orders of the instanton expansion must be resummed near the phase transition. The Hastings-McLeod solution also provides a clear example of the Lee-Yang phenomenon: the phase transition of the Gross-Witten-Wadia model appears as the $N\to\infty$ coalescence of complex zeros of the finite $N$ partition function, since in the double-scaling limit this is described by the Hastings-McLeod solution, which has singularities only in two wedges of the complex plane, pinching the real axis at the phase transition at $x=0$ \cite{novokshenov,huang,dunne}.

In this paper we study the resurgent trans-series structure of a class of non-integrable deformations of the Painlev\'e II equation, in which we deform the power of the non-linear term away from the integrable value of 3:
\begin{eqnarray}
y^{\prime\prime}(x)=2 y^N(x)+x\, y(x) \qquad, \quad N\in {\mathbb Z^+}
\label{eq:pN}
\end{eqnarray}
We show that for each $N\geq 2$, this equation  has a unique real solution, which we denote as $y_N(x)$, which is  analogous to the Hastings-McLeod solution satisfying analogous boundary conditions at $x=\pm\infty$:
\begin{eqnarray}
%\lim_{x\to\infty} 
y_N(x) &\sim& k_N {\rm Ai}(x)  \qquad, \qquad x\to+\infty
\label{eq:pNbc1} \\
%\lim_{x\to-\infty}
 y_N(x) &\sim& \left(\frac{-x}{2}\right)^{\frac{1}{N-1}}  \qquad, \quad x\to-\infty
\label{eq:pNbc2}
\end{eqnarray}
Here $k_N$ is a real constant whose value depends on the integer-valued non-linearity parameter $N$. 

When $N=3$, equations (\ref{eq:pN}-\ref{eq:pNbc2})  reduce to the Hastings-McLeod case for the Painlev\'e II equation, with $k_3=1$, which is integrable: in the vicinity of a moveable singularity (i.e. one related to the boundary conditions) at $x_0$, the solution can be expanded with only pole singularities \cite{ince}:
\begin{eqnarray}
y_3(x)=\frac{1}{x-x_0}-\frac{x_0}{6}(x-x_0)-\frac{1}{4}(x-x_0)^2+h_0(x-x_0)^3+\frac{x_0}{72}(x-x_0)^4+\dots
\label{eq:p2poles}
\end{eqnarray}
This solution is completely characterized by the pole location $x_0$ and the coefficient $h_0$ of the $(x-x_0)^3$ term: all other expansion coefficients are expressed as  polynomials in $x_0$ and $h_0$. For $N\neq 3$ such an expansion is not possible. In this paper we study the effect of this non-integrability on the resurgent trans-series structure of the solutions.

\section{Numerical Computation of the $k_N$ Parameter}
\label{sec:kN}

\begin{figure}[htb]
    \centering
    \begin{minipage}{0.45\textwidth}
        \centering
        \includegraphics[width=0.9\textwidth]{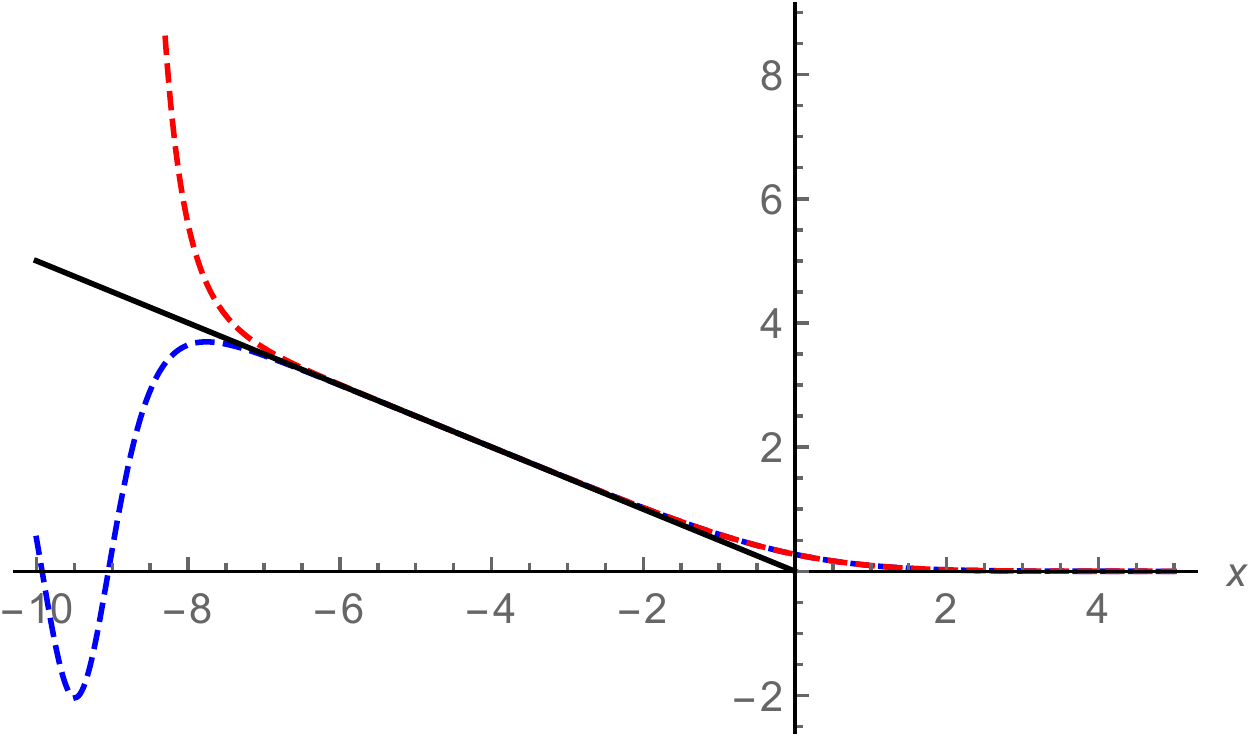} % first figure itself
        \caption{Numerical solution to (\ref{eq:pN}) with $N = 2$, $k_2 \approx 0.671231$. The red dashed and blue dotted curves correspond to $k_2=0.671232$ and $k_2=0.671231$, respectively. The solid black line is the $x<0$ asymptotic behavior in (\ref{eq:pNbc2}).}
        \label{fig:k2}
    \end{minipage}\hfill
    \begin{minipage}{0.45\textwidth}
        \centering
        \includegraphics[width=0.9\textwidth]{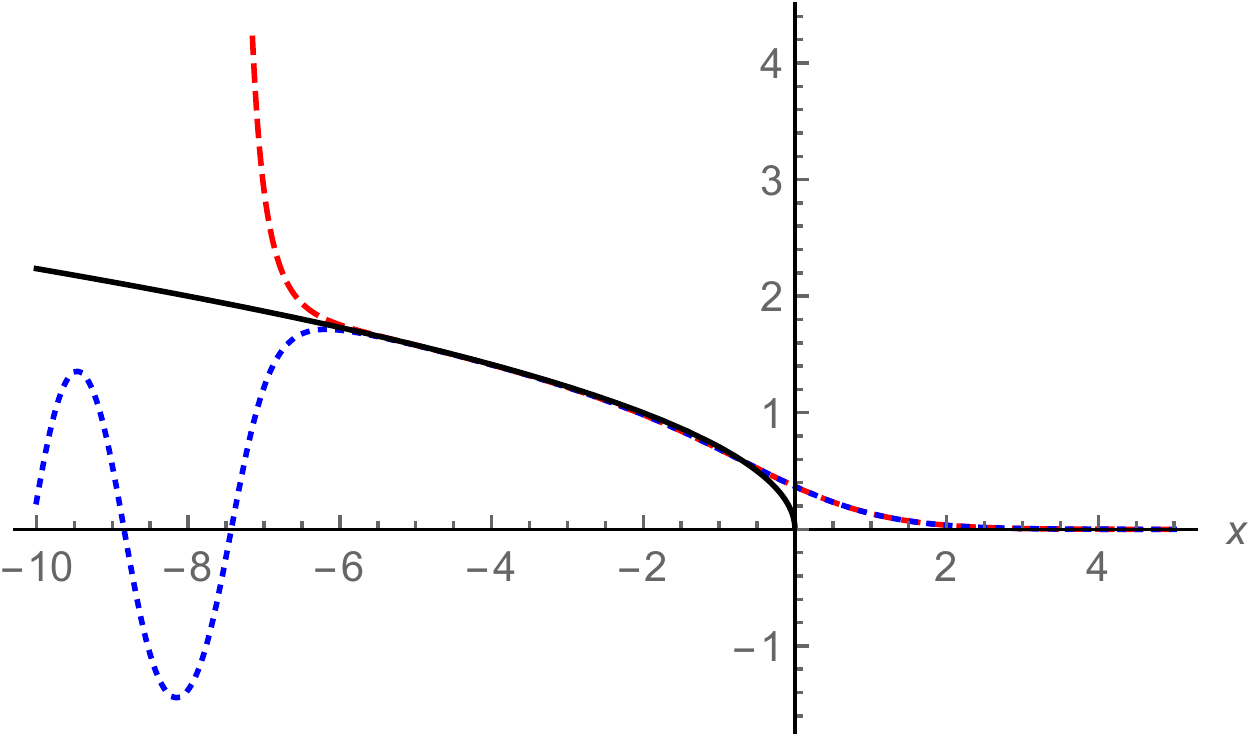} % second figure itself
        \caption{Numerical solution to (\ref{eq:pN}) with $N = 3$, $k_3 = 1$. This is the Hastings-McLeod Painlev\'e II case \cite{rosales,hastings}. The red dashed and blue dotted curves correspond to $k_3=1.00000001$ and $k_3=.9999999$, respectively. The solid black line is the $x<0$ asymptotic behavior in (\ref{eq:pNbc2}).}
        \label{fig:k3}
    \end{minipage}
%\end{figure}
%\begin{figure}[htb]
%    \centering
    \begin{minipage}{0.45\textwidth}
        \centering
        \includegraphics[width=0.9\textwidth]{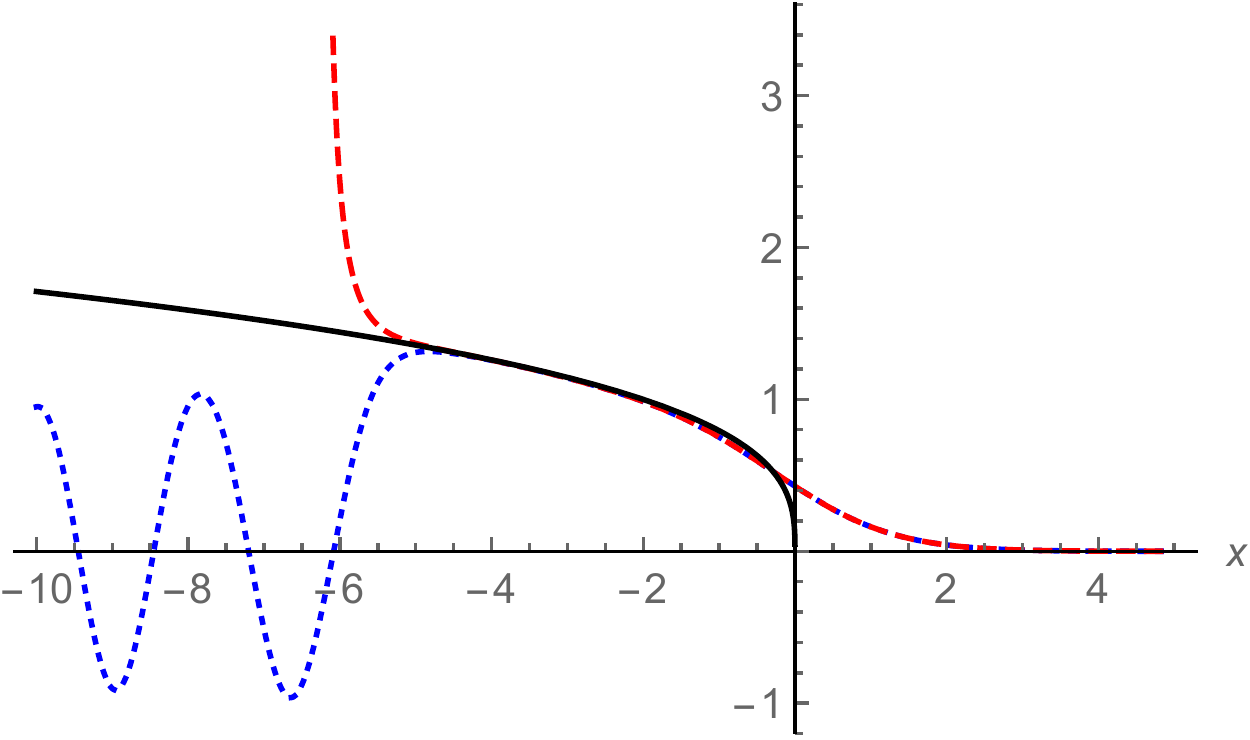} % first figure itself
        \caption{Numerical solution to (\ref{eq:pN}) with $N = 4$, $k_4 \approx 1.191124$. The red dashed and blue dotted curves correspond to $k_4=1.191125$ and $k_4=1.191124$, respectively. The solid black line is the $x<0$ asymptotic behavior in (\ref{eq:pNbc2}).}
        \label{fig:k4}
    \end{minipage}\hfill
    \begin{minipage}{0.45\textwidth}
        \centering
        \includegraphics[width=0.9\textwidth]{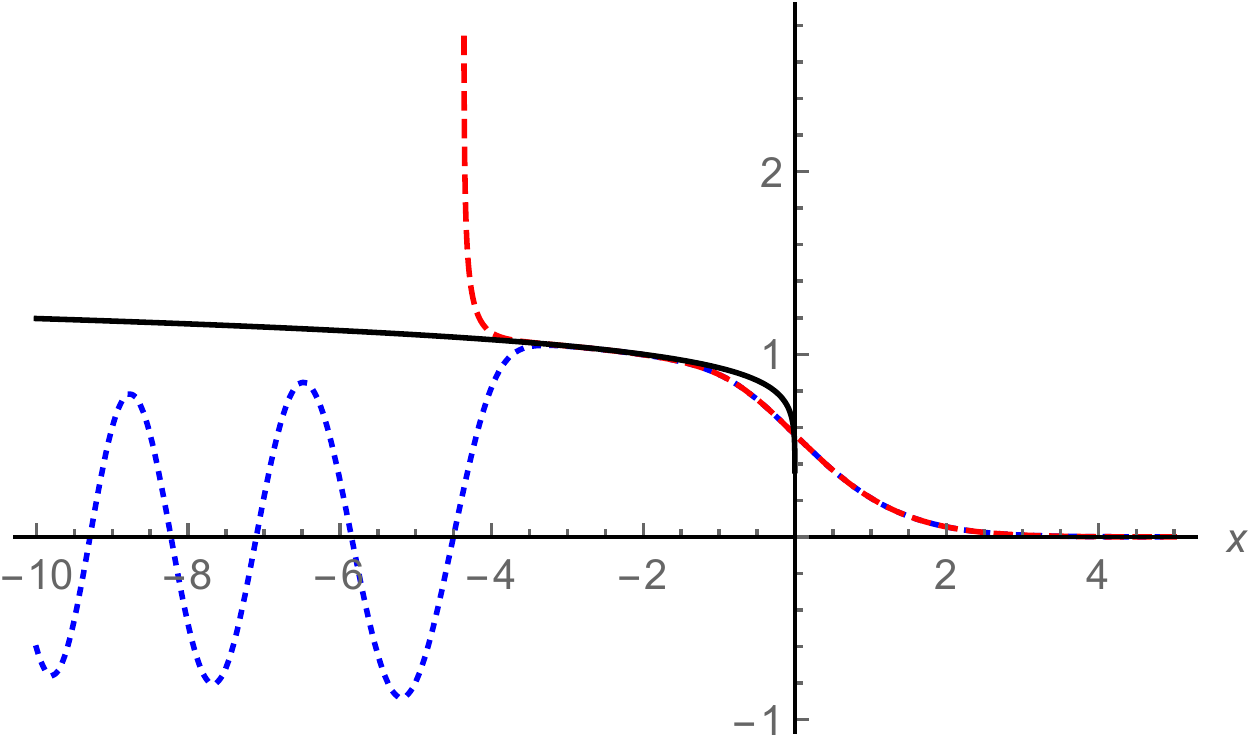} % second figure itself
        \caption{Numerical solution to (\ref{eq:pN}) with $N = 10$, $k_{10} \approx 1.577094$. The red dashed and blue dotted curves correspond to $k_{10}=1.577095$ and $k_{10}=1.577094$, respectively. The solid black line is the $x<0$ asymptotic behavior in (\ref{eq:pNbc2}).}
        \label{fig:k10}
    \end{minipage}
\end{figure}

To begin, we study numerically the boundary condition parameter $k_N$ in (\ref{eq:pNbc1}), as a function of the non-linearity parameter $N$.
This computation was made using an explicit Runge-Kutta method in Mathematica 12, with 32-digit working precision. For each of the various $N$ values we use a shooting method, starting at a large positive value of $x$ with the boundary condition (\ref{eq:pNbc1}), to tune the respective $k_N$ to match the smoothly to the behavior in (\ref{eq:pNbc2}) in the $x\to-\infty$ limit. 
Just as for the Hastings-McLeod solution of Painlev\'e II, we find that there is a unique separatrix value of $k_N$ that is sensitive to the tuning such that less than one part in $10^6$ difference results in a drastic change in the asymptotic behavior in the $x \to -\infty$ regime. 
An initial condition less than the critical $k_N$  value results in  oscillation about the negative $x$ axis, while an initial condition greater than the critical value results in a divergence of $y_N$ as $x \to -\infty$. This behavior is illustrated in Figures \ref{fig:k2}-\ref{fig:k10},  for $N=2, 3, 4, 10$. Table \ref{tab:1} lists  numerical values for the critical $k_N$ for further values of $N$. 

\begin{figure}[ht]
\centering
\includegraphics[width=0.5\textwidth]{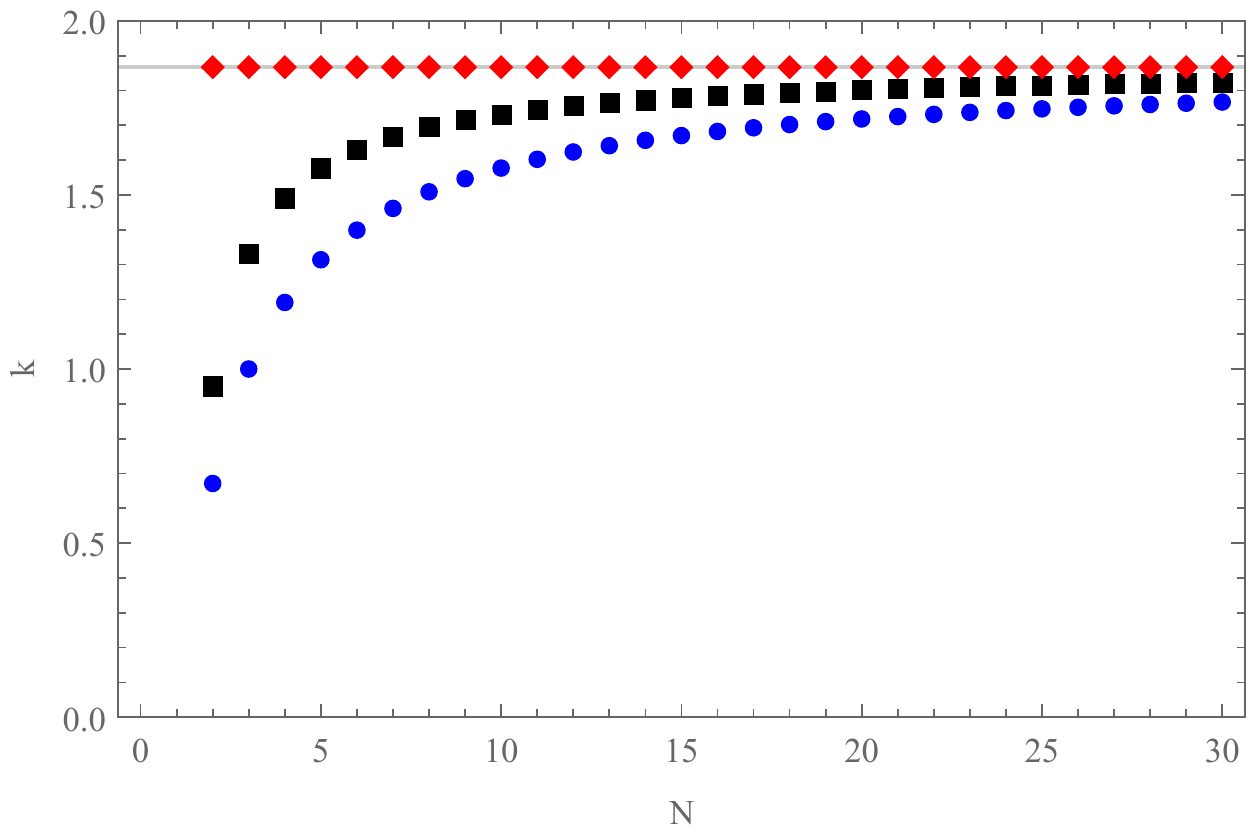}
\caption{Numerical results for the boundary condition parameter $k_N$ [blue points] as a function of the non-linearity parameter $N$ in the non-linear equation (\ref{eq:pN}). The black and red points show the upper bound, $\frac{1}{Ai(x_0)}\left(\frac{-x_0}{2}\right)^{\frac{1}{N - 1}}$, and the $N\to\infty$ upper bound, $\frac{1}{Ai(x_0)}$,  in  (\ref{eq:kNbound}) and   (\ref{eq:limit}) respectively.  }
\label{fig:kNvsN}
\end{figure}

We obtain some analytic understanding of the $N$ dependence of $k_N$ from the following argument. As the numerical solution with initial condition (\ref{eq:pNbc1}) enters the negative $x$ regime, the Airy function behavior continues in the form of an oscillatory solution if $k_N$ is less than the critical separatrix value. This means that there is a first maximum of this numerically integrated solution. For all $N$ we observe that this first maximum lies below the eventual asymptotic form, $y_N(x)\sim\left(\frac{-x}{2}\right)^{\frac{1}{N-1}}$, so we obtain an upper bound 
\begin{eqnarray}
k_N \leq \frac{1}{{\rm Ai}(x_0)}\left(\frac{-x_0}{2}\right)^{\frac{1}{N-1}} 
\label{eq:kNbound}
\end{eqnarray}
where $x_0 = -1.018792972...$ is the first zero of ${\rm Ai}'(x)$. 
In the $N\to\infty$ limit this implies an estimate
\begin{eqnarray}
\lim_{N\to\infty} k_N = \frac{1}{{\rm Ai}(x_0)} \approx 1.866867495...
\label{eq:limit}
\end{eqnarray}
This bound is supported by the large $N$ values listed in Table \ref{tab:1} and plotted in Fig. \ref{fig:kNvsN}.

\begin{table}[!htb]
    \caption{Numerically computed $k_N$ values for selected $N$ values. For each $N$, $k_N$ is the boundary condition parameter  such that both  (\ref{eq:pNbc1}) and  (\ref{eq:pNbc2}) are satisfied.}
    \begin{minipage}{.5\linewidth}
      \centering
        \begin{tabular}{l|l}
          N & $k_N$ \\\hline
    2 & 0.671231\\
    3 & 1.000000\\
    4 & 1.191124\\
    5 & 1.313788\\
    6 & 1.398870\\
    7 & 1.461275\\
    8 & 1.508981\\
    9 & 1.546629\\
    10& 1.577094\\
    11& 1.602253\\
    12& 1.623380\\
    13& 1.641374\\
    14& 1.656884\\
    15& 1.670391\\
    16& 1.682259\\
    17& 1.692771\\
    18& 1.702147\\
        \end{tabular}
    \end{minipage}%
    \begin{minipage}{.5\linewidth}
      \centering

        \begin{tabular}{l|l}
        N & $k_N$ \\\hline
    19& 1.710561\\
    20& 1.718154\\
    21& 1.725042\\
    22& 1.731317\\
    23& 1.737059\\
    24& 1.742333\\
    25& 1.747193\\
    26& 1.751687\\
    27& 1.755855\\
    28& 1.759730\\
    29& 1.763344\\
    30& 1.766720\\
    50& 1.806210\\
    100 &1.836265\\
    500 &1.860675\\
    1000 &1.863762\\
    10000 &1.866555\\
        \end{tabular}
    \end{minipage} 
    \label{tab:1}
\end{table}

A complementary approach to the separatrix parameter $k_N$ is to recast the differential equation (\ref{eq:pN}-\ref{eq:pNbc2}) as an integral equation in the $x\geq 0$ region:
\begin{eqnarray}
y(x)=k_N\, {\rm Ai}(x)+2\pi \int_x^\infty dz\, y^N(z)\left[{\rm Ai}(x) \, {\rm Bi}(z)-{\rm Ai}(z) \, {\rm Bi}(x)\right]
\label{eq:intN}
\end{eqnarray}
It is straightforward to verify that this is equivalent to the differential equation (\ref{eq:pN}), with boundary condition (\ref{eq:pNbc1}).
The integral equation (\ref{eq:intN}) can be iterated, resulting in an expansion of $y(x)$ in powers of $k_N$. Once again we find that the parameter $k_N$ must be tuned with high precision in order to match the smooth non-oscillatory behavior in (\ref{eq:pNbc2}) as $x\to-\infty$.

\section{Trans-series Solution for Positive $x$}
\label{sec:positive}

In the $x\to+\infty$ region, the boundary condition in (\ref{eq:pNbc1}) has the familiar asymptotic expansion of the Airy function:
\begin{eqnarray}
y_N(x)\sim k_N\,  \frac{e^{-\frac{2}{3}\, x^{3/2}}}{2\,\sqrt{\pi}\, x^{1/4}}\sum_{m=0}^\infty \frac{(-1)^m \Gamma\left(m+\frac{1}{6}\right)\, \Gamma\left(m+\frac{5}{6}\right)}{2\pi\,\left(\frac{4}{3} x^{3/2}\right)^m\, m! } \qquad, \quad x\to +\infty
\label{eq:pNtrans0}
\end{eqnarray}
For the full nonlinear equation (\ref{eq:pN}), this expansion generalizes to a trans-series expansion of the form
\begin{eqnarray}
y_N(x)\sim \sum_{n=0}^\infty \left(\frac{k_N\, e^{-\frac{2}{3}\, x^{3/2}}}{2\,\sqrt{\pi}\, x^{1/4}}\right)^{(N-1)n+1} {\mathcal F}_{(n)}(x)
\label{eq:pNtrans}
\end{eqnarray}
where specific powers (depending on the parameter $N$) of the Airy exponential factor are multiplied by a fluctuation series of the form
\begin{eqnarray}
{\mathcal F}_{(n)}(x)\sim \frac{1}{x^n} \sum_{m=0}^\infty \frac{d_m^{(n)}}{x^{3m/2}}
\label{eq:pNflucs}
\end{eqnarray}
The trans-series expansion in (\ref{eq:pNtrans}) has the characteristic structure of a ``multi-instanton'' expansion in physics, as studied for the 
Painlev\'e II case ($N=3$) in \cite{marino,ahmed} for applications to the strong-coupling region of the Gross-Witten-Wadia unitary matrix model. This identifies the boundary condition parameter $k_N$ as the trans-series parameter \cite{costin,marino,ahmed}, which must be tuned to a special unique value in order for this trans-series to match smoothly across the phase transition at $x=0$ to the $x\to-\infty$ behavior in (\ref{eq:pNbc2}).

Inserting the trans-series ansatz (\ref{eq:pNtrans}-\ref{eq:pNflucs}) into the differential equation (\ref{eq:pN}), we obtain a tower of non-linear recursion formulas for the fluctuation coefficients $d_m^{(n)}$, one for each non-perturbative sector labeled by the index $n$. For example, in the $n=1$ sector, the first non-trivial ``instanton'' sector, we find the recursion relation for the $d_m^{(1)}$ fluctuation coefficients (for $m=0,1,2, \dots$): 
\begin{eqnarray}
d^{(1)}_m +
     \frac{N}{2}(N + 6 m - 3) d^{(1)}_{m-1} + 
    \frac{1}{16} (N + 6 m - 8) (N + 6 m - 4)d^{(1)}_{m-2}
= \frac{b_{m + 1}}{(N^2 - 1)}      %\qquad ,\quad m=0, 1, 2, \dots
     \label{eq:d1m-rec}
     \end{eqnarray}
with initial conditions $d^{(1)}_{-2}=d^{(1)}_{-1}=0$. Here the $b_{m}$  are the expansion coefficients obtained by raising the Airy function asymptotic expansion to the $N^{\rm th}$ power:
\begin{eqnarray}
2\left(\sum_{l=0}^\infty \frac{\left(-1\right)^l \Gamma\left(l+\frac{1}{6}\right) 
\Gamma\left(l+\frac{5}{6}\right)}{2\pi \left(\frac{4}{3}\, x^{3/2}\right)^l\, l!} \right)^N \equiv \sum_{m=0}^\infty \frac{b_m}{x^{3m/2}}
\label{eq:bm}
\end{eqnarray}
%\begin{eqnarray}
%2\left(\sum_{l=0}^\infty \frac{\left(-\frac{3}{2}\right)^l \Gamma\left(3l+\frac{1}{2}\right)\, z^l}{54^l\, \Gamma\left(l+\frac{1}{2}\right)\,\Gamma\left(l+1\right)} \right)^N \equiv \sum_{m=0}^\infty b_m z^m
%\label{eq:bm}
%\end{eqnarray}
The recursion relation (\ref{eq:d1m-rec}) can be used to generate the ``one-instanton'' fluctuation coefficients $d_m^{(1)}$ to very high order, for various values of the non-linearity parameter $N$. For all $N$, these coefficients alternate in sign and grow factorially in magnitude. Using the generated coefficients, combined with Richardson acceleration \cite{bender-book}, we have found the following results for the leading and sub-leading asymptotics:
\begin{eqnarray}
%N=2\,:\, d_m^{(1)}& \sim& -(-1)^m \left(\frac{3}{2}\right)^m \left(m-\frac{1}{6}\right)!\left[1+\frac{\frac{2}{3}\cdot \frac{5}{48}}{\left(m-\frac{1}{6}\right)}+\dots \right] \quad , \quad m\to\infty\\
%N=3\,:\, d_m^{(1)}& \sim& (...)^m \,\\
N=4\,:\, d_m^{(1)}& \sim& (-1)^{m} \frac{1}{\pi} 
\left(\frac{3}{4}\right)^{m-1} \Gamma(m)\left[1+ \frac{\frac{61}{18}}{(m-1)}+\dots \right] \quad , \quad m\to\infty
\label{eq:dm-large1}\\
N=5\,:\, d_m^{(1)}& \sim& (-1)^{m}  \frac{15}{32\pi}
\left(\frac{3}{4}\right)^{m-1} \Gamma(m) \left[1+\frac{\frac{29}{12}}{(m-1)}+\dots \right] \quad , \quad m\to\infty
\label{eq:dm-large2}\\
N=6\,:\, d_m^{(1)}& \sim& (-1)^{m} \frac{3}{10\pi}
\left(\frac{3}{4}\right)^{m-1} \Gamma(m)\left[1+\frac{\frac{97}{45}}{(m-1)}+\dots \right] \quad , \quad m\to\infty
\label{eq:dm-large3}\\
N=7\,:\,  d_m^{(1)}& \sim& (-1)^{m}  \frac{7}{32\pi}
\left(\frac{3}{4}\right)^{m-1} \Gamma(m)\left[1+\frac{ \frac{25}{12}}{(m-1)}+\dots \right] \quad , \quad m\to\infty
%N=4\,:\, d_m^{(1)}& \sim& (-1)^{m+1} \left(\frac{3}{4}\right)^m m!\left[1+\frac{\frac{4}{3}\cdot \frac{61}{32}}{m}+\dots \right] \quad , \quad m\to\infty\\
%N=5\,:\, d_m^{(1)}& \sim& (-1)^{m+1} \left(\frac{3}{4}\right)^m m!\left[1+\frac{\frac{4}{3}\cdot \frac{87}{64}}{m}+\dots \right] \quad , \quad m\to\infty\\
%N=6\,:\, d_m^{(1)}& \sim& (-1)^{m+1} \left(\frac{3}{4}\right)^m m!\left[1+\frac{\frac{4}{3}\cdot \frac{97}{90}}{m}+\dots \right] \quad , \quad m\to\infty\\
%N=7\,:\,  d_m^{(1)}& \sim& (-1)^{m+1}  \left(\frac{3}{4}\right)^m m!\left[1+\frac{\frac{4}{3}\cdot \frac{75}{64}}{m}+\dots \right] \quad , \quad m\to\infty
%\\
\label{eq:dm-large4}
\end{eqnarray}
These expansions can also be confirmed using the integral equation (\ref{eq:intN}). Inserting a trans-series ansatz of the form
\begin{eqnarray}
y(x)\sim \sum_{n=0}^\infty k_N^{(N-1)n+1} \, Y_{[(N-1)n+1]}(x)\qquad, \quad x\to+\infty
\label{eq:pos-intN}
\end{eqnarray}
expanded in powers of the trans-series parameter $k_N$, 
yields a tower of {\it linear} equations with solutions:
\begin{eqnarray}
Y_{[1]}(x)&=& {\rm Ai}(x) \\
Y_{[N]}(x)&=& 2\pi \left({\rm Ai}(x) \int_x^\infty \left(Y_{[1]}(z)\right)^N  {\rm Bi}(z)\, dz -{\rm Bi}(x) \int_x^\infty \left(Y_{[1]}(z)\right)^N{\rm Ai}(z) \, dz\right)\\
Y_{[2N-1]}(x)&=& 2N\pi \left({\rm Ai}(x) \int_x^\infty Y_{[N]}(z) \left(Y_{[1]}(z)\right)^{N-1} {\rm Bi}(z)\, dz \right.
\nonumber\\
&&\left. \qquad 
-{\rm Bi}(x) \int_x^\infty Y_N^{(N)}(z)  \left(Y_N^{(1)}(z)\right)^{N-1} {\rm Ai}(z) \, dz\right)\label{eq:int-sols}
\\
&\vdots & 
\nonumber
\end{eqnarray}
The corresponding asymptotic expansions can therefore be generated  from those of the Airy functions. This trans-series structure generalizes that of the $N=3$ case, the Painlev\'e II equation \cite{ahmed}. The $N=2$ case is special: see Section \ref{sec:N2} below.

\section{Trans-series Solution for Negative $x$}
\label{sec:negative}

In the $x\to -\infty$ limit the trans-series has a completely different structure.  As $x\to-\infty$, the boundary condition in (\ref{eq:pNbc2}) naively generalizes to a formal series expansion of the form
\begin{eqnarray}
y_N(x)\sim \left(\frac{-x}{2}\right)^{\frac{1}{N-1}}\sum_{l=0}^\infty \frac{c_l}{(-x)^{3l}}\qquad, \quad x\to -\infty
\label{eq:pNtrans2}
\end{eqnarray}
Here $c_0\equiv 1$. The higher expansion coefficients $c_l$ can be generated recursively by inserting this ansatz into the differential equation (\ref{eq:pN}). For example, we find
\begin{eqnarray}
 c_1=\frac{(N-2)}{(N-1)^3}
\label{eq:cls}
\end{eqnarray}
[The $N=2$ case is special, as discussed below in Section \ref{sec:N2}.]
The remaining terms can be generated from a recursion relation. First note that for the expansion (\ref{eq:pNtrans2}), we find that 
\begin{eqnarray}
\frac{d^2 y_N}{dx^2}\sim \frac{2}{(N-1)^2} \left(\frac{-x}{2}\right)^{\frac{N}{N-1}}\sum_{l=1}^\infty \frac{\left[(3l-3)N-(3l-2))((3l-2)N-(3l-1)\right] c_{l-1}}{(-x)^{3l}}
\label{eq:d2yNdx2}
\end{eqnarray}
To satisfy the differential equation we see that the expansion coefficients are given by
\begin{eqnarray}
c_{l-1}=\frac{(N-1)^2}{\left[(3l-3)N-(3l-2))((3l-2)N-(3l-1)\right]}\,g_l \quad, \quad l=1, 2, 3, \dots
\label{eq:neg-rec}
\end{eqnarray}
where the $g_l$ are the coefficients of the expansion:
\begin{eqnarray}
 \left(\sum_{l=0}^\infty \frac{c_l}{(-x)^{3l}}\right)^N-\left(\sum_{l=0}^\infty \frac{c_l}{(-x)^{3l}}\right) \equiv \sum_{l=1}^\infty \frac{g_l}{(-x)^{3l}}
\label{eq:gl}
\end{eqnarray}
However, it is clear that the expansion (\ref{eq:pNtrans2}) cannot be the full expansion, because there is no boundary condition parameter: all the $c_l$ coefficients are completely determined recursively. The ``missing'' boundary condition parameter enters because the expansion (\ref{eq:pNtrans2}) is an asymptotic formal series which must be completed with an infinite sum of exponentially small non-perturbative terms, resulting in the trans-series expansion\footnote{Note that 
for $n\geq 1$, the fluctuation expansions are actually in powers of $1/(-x)^{3/2}$ rather than $1/(-x)^{3}$.}
\begin{eqnarray}
y_N(x)\sim \left(\frac{-x}{2}\right)^{\frac{1}{N-1}}\sum_{n=0}^\infty  \left(\sigma_N \frac{e^{-  \frac{2}{3} \sqrt{N-1} (-x)^{3/2}}}{(-x)^{1/4}}\right)^n \sum_{l=0}^\infty \frac{c^{(n)}_l}{(-x)^{3l/2}}\qquad, \quad x\to -\infty
\label{eq:pNtrans3}
\end{eqnarray}
The trans-series parameter $\sigma_N$ characterizes this family of asymptotic solutions, all having the same leading behavior (\ref{eq:pNbc2}) as $x\to-\infty$. 
This $x\to -\infty$ trans-series parameter  $\sigma_N$ must be simultaneously tuned, together with the trans-series parameter $k_N$ in the $x\to+\infty$ trans-series (\ref{eq:pNtrans}), in order to obtain the unique real solution matching {\it both} the $x\to\pm\infty$ boundary conditions in (\ref{eq:pNbc1}-\ref{eq:pNbc2}). 

Notice that the exponent of the instanton factor in (\ref{eq:pNtrans3}) differs from that in the $x\to+\infty$ trans-series (\ref{eq:pNtrans}) by more than just $x\to-x$: there is an additional factor of $\sqrt{N-1}$. Physically, this means that as the phase transition (at $x=0$) is approached from either side, all instanton orders of the appropriate trans-series are required, and a highly non-trivial instanton condensation phenomenon occurs, resumming all instanton terms into a different instanton factor, as has been studied for the Painlev\'e II equation \cite{bothner}. This is an example of the {\it nonlinear} Stokes phenomenon, characteristic of nonlinear ODEs, and very different from the familiar linear Stokes phenomenon.
The instanton exponent  in (\ref{eq:pNtrans3}) can be deduced in two complementary ways. First, we  linearize the equation,
 writing $y_N(x)$ as a sum of a perturbative piece plus an exponentially small correction, $y_N=y_{\rm pert}+\delta y$,
 from which  the leading $x\to -\infty$ behavior follows:
\begin{eqnarray}
\delta y^{\prime\prime}\sim -(N-1)x\, \delta y \quad \Rightarrow \quad \delta y\sim e^{-  \frac{2}{3} \sqrt{N-1} (-x)^{3/2}}
\label{eq:dy}
\end{eqnarray}
Second, the exponential factor can be deduced from a study of the large order behavior of the perturbative expansion coefficients $c_l$ in the expansion (\ref{eq:pNtrans2}). The expansion coefficients are determined from the recursion formula (\ref{eq:neg-rec}), and we find the large-order behavior:
\begin{eqnarray}
%N=2\,:\, c_l^{(0)}& \sim& 0 \quad , \quad l\to\infty\\
N=3\,:\, c_l & \sim&  -(0.1466323...) \times \left(\frac{9}{8}\right)^l \Gamma\left(2l-\frac{1}{2}\right) \quad , \quad l\to\infty \\
N=4\,:\, c_l & \sim& -(0.12985376...) \times \left(\frac{9}{12}\right)^l \Gamma\left(2l-\frac{7}{18}\right) \quad , \quad l\to\infty \\
N=5\,:\, c_l & \sim&  -(0.1086460...) \times \left(\frac{9}{16}\right)^l \Gamma\left(2l-\frac{1}{3}\right) \quad , \quad l\to\infty 
\label{eq:cl-large}
\end{eqnarray}
The factorial growth goes like $(2l)!$, and the multiplicative power is $\left(\frac{3}{2\sqrt{N-1}}\right)^{2l}$. A general Borel summation argument implies that the corresponding exponential factor should be the square root (since the growth is $(2l)!$ rather than $l!$) of the inverse of $\left(\frac{3}{2\sqrt{N-1}}\right)^2$, as in (\ref{eq:pNtrans3}) and (\ref{eq:dy}).

Note that the perturbative coefficients $c_l$ diverge factorially in magnitude (for $N\geq 3$), and do not alternate in sign. Therefore, if we wish the trans-series expansion (\ref{eq:pNtrans3}) to represent a real solution to the equation, the trans-series parameter $\sigma_N$ must have an imaginary part, in order for the higher-order exponential terms to cancel the imaginary terms generated by Borel summation of the factorially divergent non-sign-alternating perturbative expansion \cite{ines}. This mimics the well-known behavior of the Painlev\'e II trans-series in the $x\to -\infty$ region \cite{costin,marino}.

\section{Trans-series for the $N=2$ case}
\label{sec:N2}

The $N=2$ case is interestingly different from the other cases, for $N\geq 3$. The large-order behavior of the first fluctuation terms in the $x\to+\infty$ region displays a different growth rate [contrast with (\ref{eq:dm-large1})-(\ref{eq:dm-large4})]:
\begin{eqnarray}
N=2\,:\, d_m^{(1)}& \sim& (4.9161362...)  \times  (-1)^{m} \left(\frac{3}{2}\right)^{m-1} \left(m-\frac{1}{6}\right)!
%\left[1+\frac{\frac{2}{3}\cdot \frac{5}{48}}{\left(m-\frac{1}{6}\right)}+\dots \right] 
\quad , \quad m\to\infty
%N=3\,:\, d_m^{(1)}& \sim& (...)^m \,\\
%N=4\,:\, d_m^{(1)}& \sim& (-1)^{m+1} \left(\frac{3}{4}\right)^m m!\left[1+\frac{\frac{4}{3}\cdot \frac{61}{32}}{m}+\dots \right] \quad , \quad m\to\infty\\
%N=5\,:\, d_m^{(1)}& \sim& (-1)^{m+1} \left(\frac{3}{4}\right)^m m!\left[1+\frac{\frac{4}{3}\cdot \frac{87}{64}}{m}+\dots \right] \quad , \quad m\to\infty\\
%N=6\,:\, d_m^{(1)}& \sim& (-1)^{m+1} \left(\frac{3}{4}\right)^m m!\left[1+\frac{\frac{4}{3}\cdot \frac{97}{90}}{m}+\dots \right] \quad , \quad m\to\infty\\
%N=7\,:\,  d_m^{(1)}& \sim& (-1)^{m+1}  \left(\frac{3}{4}\right)^m m!\left[1+\frac{\frac{4}{3}\cdot \frac{75}{64}}{m}+\dots \right] \quad , \quad m\to\infty
%\\
\label{eq:dm-largeN2}
\end{eqnarray}
However, this leads to a qualitatively similar trans-series structure since the coefficients diverge factorially and alternate in sign. The biggest difference arises in the $x\to-\infty$ region, where the leading asymptotic form in (\ref{eq:pNbc2}), $y_2(x)\sim -\frac{x}{2}$, is in fact an {\it exact} solution of the nonlinear differential equation (\ref{eq:pN}) when $N=2$. Thus, all the coefficients $c_l$ in the expansion (\ref{eq:pNtrans2}) vanish for $l\geq 1$. This is consistent with (\ref{eq:cls}). This fact has important consequences for the matched solution. Even though $y_2(x)= -\frac{x}{2}$ is an exact solution, it does not match smoothly with the $x\to +\infty$ asymptotics in (\ref{eq:pNbc1}). 
There is {\it another} solution, which has an infinite tower of exponentially small terms in the negative $x$ region, and it is this solution that matches smoothly to the Airy-like solution in the positive $x$ region. This is the $N=2$ analog of the $N=3$ Hastings-McLeod behavior. To see this, it is simplest to study the iterated integral equation form of the problem. 

For the $x\to+\infty$ region this is given by the trans-series expansion in (\ref{eq:pos-intN}-\ref{eq:int-sols}) for $N=2$:
\begin{eqnarray}
y(x)\sim k_2 Y_{[1]}(x)+  k_2^2 Y_{[2]}(x)+  k_2^3 Y_{[3]}(x) +O(k_2^4)\quad, \quad x\to+\infty
\label{eq:N2pos}
\end{eqnarray}
where
\begin{eqnarray}
Y_{[1]}(x)&=&{\rm Ai}(x)\\
Y_{[2]}(x)&=& 2\pi \left({\rm Ai}(x) \int_{x}^\infty {\rm Ai}^2(z) {\rm Bi}(z)\, dz -{\rm Bi}(x) \int_{x}^\infty {\rm Ai}^3(z) \, dz\right)
%y_2(x)\sim k_2 {\rm Ai}(x)+ 2\pi k_2^2\left({\rm Ai}(x) \int_x^\infty {\rm Ai}^2(z) {\rm Bi}(z)\, dz -{\rm Bi}(x) \int_x^\infty {\rm Ai}^3(z) \, dz\right)+O(k_2^3)\quad, \quad x\to+\infty
\label{eq:N2pos2}
\end{eqnarray}
Similarly, we can write an iterated integral equation form as $x\to-\infty$:
\begin{eqnarray}
y(x)\sim -\frac{x}{2}+\sigma_2 W_{[1]}(x)+\sigma_2^2 W_{[2]}(x)++\sigma_2^3 W_{[3]}(x)+O(\sigma_2^4)\quad, \quad x\to-\infty
\label{eq:N2neg}
\end{eqnarray}
where 
\begin{eqnarray}
W_{[1]}(x)&=&{\rm Ai}(-x)\\
W_{[2]}(x)&=& 2\pi \left({\rm Ai}(-x) \int_{-x}^\infty {\rm Ai}^2(z) {\rm Bi}(z)\, dz -{\rm Bi}(-x) \int_{-x}^\infty {\rm Ai}^3(z) \, dz\right)
%\\&\vdots& \nonumber
\label{eq:N2negW}
\end{eqnarray}
\begin{figure}[ht]
\centering
\includegraphics[width=0.5\textwidth]{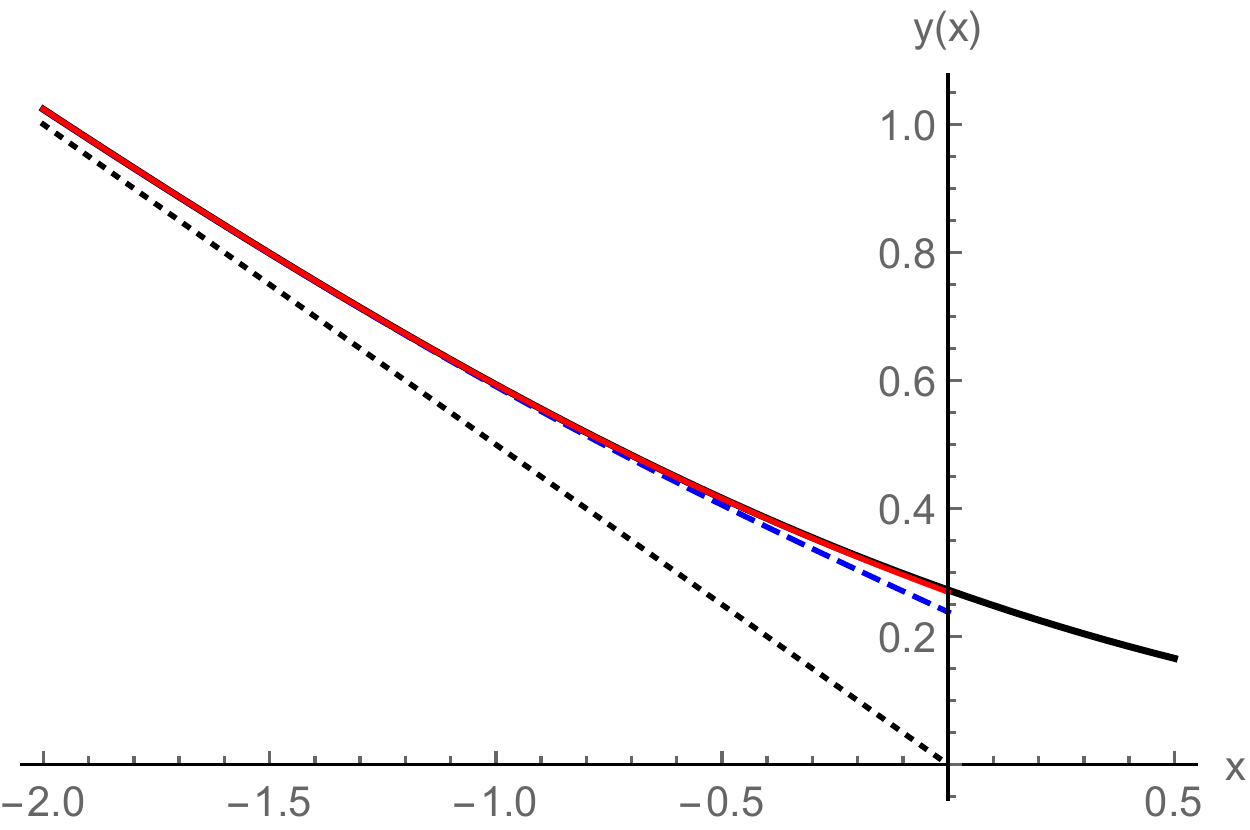}
\caption{The dashed curve is the $x\to-\infty$ asymptotics $y(x)\sim -\frac{x}{2}$, which is an exact solution  for $N=2$. The black curve is the exact numerical solution to  (\ref{eq:pN})-(\ref{eq:pNbc2}) for $N=2$. The  blue-dashed and red curves denote the $x<0$ trans-series solution in (\ref{eq:N2neg}), including the first two non-trivial trans-series contributions, $W_{[1]}(x)$ and $W_{[2]}(x)$, respectively, and with trans-series parameter $\sigma_2=k_2=0.671231$. This agrees very well with the positive $x$ solution.}
\label{fig:N2}
\end{figure}
The higher terms in the $x\to\pm \infty$ expansions  are identical, up to $x\to -x$. 
This correspondence between the expansions in (\ref{eq:N2pos})-(\ref{eq:N2pos2}) and (\ref{eq:N2neg})-(\ref{eq:N2negW}) arises because if we define $w(x)=y(x)+\frac{x}{2}$, then $w(x)$ satisfies the same equation, but with $x\to -x$: $w''=2w^2-x\, w$.
Therefore, by matching the two expansions at $x=0$, we see that we require the two trans-series parameters to be equal: $k_2=\sigma_2$. Using the numerically computed value of $k_2=0.671231...$, we plot in Figure \ref{fig:N2} the two trans-series expressions developed at $x=\pm \infty$. We observe that even with just the first exponentially suppressed terms added the agreement is extremely good, improving even further with the next term. Thus we see that even though $y_2(x)=-\frac{x}{2}$ is an {\it exact} solution, there is a non-perturbative completion, in the form of the trans-series (\ref{eq:N2neg}), which is the function that smoothly connects with the Airy-like behavior as $x\to+\infty$. This is a further indication of the exponential sensitivity of the generalized Hastings-McLeod solution in (\ref{eq:pN})-(\ref{eq:pNbc2}).

\section{Conclusions}
\label{sec:conclusions}

We have shown that the non-integrable generalizations (\ref{eq:pN}) of the Painlev\'e II equation, with boundary conditions (\ref{eq:pNbc1})-(\ref{eq:pNbc2}), have unique real solutions with many qualitative features in common with the  physical Hastings-McLeod solution of Painlev\'e II. For each power $N$, there is a unique real boundary condition parameter $k_N$ in (\ref{eq:pNbc1}) such that the $x\to+\infty$ behavior matches smoothly to the $x\to-\infty$ behavior. We find the full trans-series structure of these solutions in both asymptotic regions, indicating a nonlinear Stokes transition between the two regimes. These results demonstrate explicitly that the integrability of the Painlev\'e equations is not essential for many of the well-known results regarding trans-series and resurgent asymptotics for nonlinear differential equations.

\section{Acknowledgements}
This material is based upon work supported by the U.S. Department of Energy, Office of Science, Office of High Energy Physics under Award Number DE-SC0010339 (GD).  This work was part of an undergraduate research project (NC).

\end{document}